\documentclass[intlimits,twoside,a4paper]{article}

\usepackage{amsmath,amssymb}
\usepackage{graphicx}
\usepackage{wrapfig}
\usepackage{xcolor}

\usepackage[T2A]{fontenc}
\usepackage[cp1251]{inputenc}

\usepackage[eqsecnum]{cmpj2}

\issue{2013}{16}{2}{23005}

\doinumber{10.5488/CMP.16.23005}

\title[Energy spectrum and phase diagrams two-sublattice hard-core boson model]
{Energy spectrum and phase diagrams of two-sublattice hard-core boson model}

\author{I.V.~Stasyuk, O.~Vorobyov}

\address{Institute for Condensed Matter Physics National
academy of Sciences of Ukraine, \\1 Sventsitskii St., 79011 Lviv,
Ukraine}

\date{Received March 14, 2013, in final form March 29, 2013}

\begin{document}

\maketitle

\begin{abstract}
The energy spectrum, spectral density and phase diagrams have been obtained
for two-sublattice hard-core boson model in frames of random phase
approximation approach. Reconstruction of boson spectrum at the  change of
temperature, chemical potential and energy difference between local positions
in sublattices is studied. The  phase diagrams illustrating the  regions of
existence of a normal phase which can be close to Mott-insulator (MI) or charge-density (CDW) phases as well as  the phase with the Bose-Einstein condensate (SF phase) are built.

\pacs 03.75.Hh, 03.75.Lm, 71.35.Lk
\end{abstract}

\section{Introduction}

Lattice Bose-gas model based on the hard-core bosons approach (the site
occupancy $n_i=0,1$) has a wide range of possible applications starting from
quantum effects in liquid He \cite{matsuda,lin}. This model was also applied
to superconducting gas of Cooper electron pairs \cite{micnas}, physical
properties of Josephson junctions \cite{czathy}, thermodynamics and energy
spectrum of crystals with ionic conductivity \cite{mahan,dulepa}. In recent
years the hard-core boson approach has gained popularity in connection with
investigations of ultra-cold atoms in optical lattices. At an arbitrary
occupation of local particle positions optical lattices are usually described
with Bose-Hubbard model (see \cite{bloch} for review). In $U \rightarrow
\infty$ limit of this model, when potential wells are extremely deep,
Bose-Hubbard model turns to hard-core boson model. In this paper we consider
this model for the lattice with non-equivalent sites, particularly in the
simplest case of two-sublattice structure. Such structures can be  easily
realized in optical lattices \cite{Aizenman}  and  are also observed in the
case of adsorption of hydrogen atoms on the  surface of metals (the  quantum
surface diffusion of protons is  described by means of Bose-Hubbard model
\cite{Puska,Brenig}). Crystal lattice is supposed to be centrosymmetrical of
cubic type. Particles have different local site energies on each of two
sublattices ($\varepsilon_{\mathrm{A}} \neq \varepsilon_{\mathrm{B}}$, where A and B are sublattice
indices). This model has been investigated in connection with thermodynamic
properties of Bose atoms in complex optical lattices \cite{iskin,hen2,pich}.

The main focus of our paper is to study the conditions of Bose-Einstein (BE)
condensation and to construct the corresponding phase diagrams.
Our goal is to investigate the energy spectrum and one-particle spectral
densities as well as the changes of their shapes as the system enters
various phases that include the phase with  BE condensate (also called
superfluid or SF) and normal phase of the so-called Mott-insulator (MI) or
charge-density wave (CDW) type. We use two-time Green's function technique
and random phase approximation (RPA). A similar approach has been used recently
in \cite{dulepa2}.

\section{Boson Green's functions and phase diagrams}

The Hamiltonian of noninteracting hard-core bosons on a lattice is as follows:
\begin{equation}
\hat{H} = - \sum_{ij} t_{ij}b_i^+b_j + (\varepsilon_0-\mu) \sum_i n_i\,,
\label{GrindEQ_ista2.1}
\end{equation}
 where $t_{ij}$ is the  boson hopping parameter and $b_i,\, b_i^+$  are Pauli operators. We proceed to pseudospins ($b_i=S_i^+$, $b_i^+=S_i$) and generalize
 the model for two sublattices ($i=n.\alpha$; $\alpha=A,B$; $\varepsilon_{0}=\varepsilon_{\mathrm{A}}$, $\varepsilon_{\mathrm{B}}$):
\begin{eqnarray}
\hat{H} = - \sum_{n\alpha} \sum_{n'\beta}J_{nn'}^{\alpha\beta}\left( S_{n\alpha}^x  S_{n'\beta}^x + S_{n\alpha}^y S_{n'\beta}^y\right) %
- \sum_{\alpha}h_{\alpha} \sum_{n}  S_{n\alpha}^z\,.
\label{GrindEQ_ista2.2}
\end{eqnarray}

The parameter of ``transversal'' interaction between pseudospins
$J_{nn'}^{\alpha\beta}$ describes the transfer of particles between  nearest
neighbours in the lattice; $h_{\alpha}=\varepsilon_{\alpha}-\mu$ is the
``field'' acting on  the pseudospin in  $\alpha$ sublattice.

To start with, we consider the  mean-field Hamiltonian
\begin{equation}
\hat{H}_{\mathrm{MF}} = - \sum_{n\alpha}\sum_{n'\beta}\left(J_{nn'}^{\alpha\beta}+J_{n'n}^{\beta\alpha}\right)\langle S_{\beta}^{x}\rangle S_{n\alpha}^{x}
-\sum_{\alpha}h_{\alpha}\sum_{n}S_{n\alpha}^{z}
\label{GrindEQ_ista2.3}
\end{equation}
which is diagonalized with the rotation transformation
\begin{eqnarray}
S_{n\alpha}^z&=&\sigma_{n\alpha}^z \cos{\vartheta_{\alpha}} + \sigma_{n\alpha}^x \sin{\vartheta_{\alpha}}\,, \nonumber \\
S_{n\alpha}^x&=&\sigma_{n\alpha}^x \cos{\vartheta_{\alpha}} - \sigma_{n\alpha}^z \sin{\vartheta_{\alpha}}
\label{GrindEQ_ista2.4}
\end{eqnarray}
and takes the  form
$\hat{H}_{\mathrm{MF}}=-\sum_{n\alpha}E_{\alpha}\sigma_{n\alpha}^{z}$.

The following equations define the angles $\vartheta_{\alpha}$:
\begin{eqnarray}
h_{\mathrm{A}}\sin\vartheta_{\mathrm{A}}-\langle\sigma_{\mathrm{B}}^{z}\rangle J_{\mathrm{A}}(0)\cos\vartheta_{\mathrm{A}}\sin\vartheta_{\mathrm{B}}=0,\nonumber\\
h_{\mathrm{B}}\sin\vartheta_{\mathrm{B}}-\langle\sigma_{\mathrm{A}}^{z}\rangle J_{\mathrm{B}}(0)\cos\vartheta_{\mathrm{B}}\sin\vartheta_{\mathrm{A}}=0.
\label{GrindEQ_ista2.5}
\end{eqnarray}
Here,
$J_{\alpha}(0)=\sum_{n'\beta}\left(J_{nn'}^{\alpha\beta}+J_{n'n}^{\beta\alpha}\right)$;
in the  case of structurally equivalent sublattices $J_{\mathrm{A}}(0)=J_{\mathrm{B}}(0)\equiv
J(0)$.

The trivial solution $\sin{\vartheta_{\mathrm{A}}}=0, \,\sin{\vartheta_{\mathrm{B}}}=0$ defines
the normal phase (like MI or CDW), while at $\sin{\vartheta_{\alpha}} \neq 0$
the SF phase exists. For SF phase, the order parameter $\langle S_{\alpha}^x
\rangle$ is not equal to zero (because $\langle S_{\alpha}^x \rangle = -
\langle \sigma_{\alpha}^x \rangle \sin{\vartheta_{\alpha}}$).

For nontrivial solution we have
\begin{equation}
\sin^{2}\vartheta_{\alpha}=\frac{\langle \sigma_{\alpha}^{z}\rangle^{2}\langle \sigma_{\beta}^{z}\rangle^{2} J^{4}(0)-h_{\alpha}^{2}h_{\beta}^{2}}
{\langle \sigma_{\alpha}^{z}\rangle^{2}J^{2}(0)[h_{\alpha}^{2}+\langle \sigma_{\beta}^{z}\rangle^{2}J^{2}(0)]}\,.
\label{GrindEQ_ista2.6}
\end{equation}
Here and below, $\beta\neq\alpha$. In the  mean-field approximation
\begin{equation}
\langle \sigma_{\alpha}^{z}\rangle=\frac12\tanh \frac{\beta E_{\alpha}}{2}\,,
\label{GrindEQ_ista2.7}
\end{equation}
where
\begin{eqnarray}
E_{\alpha} = h_{\alpha}\cos\vartheta_{\alpha}+\langle\sigma_{\beta}^{z}\rangle J(0)\sin\vartheta_{\alpha}\sin\vartheta_{\beta}
=\langle\sigma_{\alpha}^{z}\rangle J(0)\frac{\sqrt{h_{\alpha}^{2}+\langle\sigma_{\beta}^{z}\rangle^{2} J^{2}(0)}}
{\sqrt{h_{\beta}^{2}+\langle\sigma_{\alpha}^{z}\rangle^{2} J^{2}(0)}}\,.
\label{GrindEQ_ista2.8}
\end{eqnarray}
The set of equations (\ref{GrindEQ_ista2.7}) and (\ref{GrindEQ_ista2.8})
defines the pseudospin averages $\langle\sigma_{\mathrm{A}}^{z}\rangle$,
$\langle\sigma_{\mathrm{B}}^{z}\rangle$ and internal fields $E_{\mathrm{A}}$, $E_{\mathrm{B}}$. On  the
other hand, in  the case of normal phase
\begin{equation}
E_{\alpha}=h_{\alpha}\,, \qquad\langle \sigma_{\alpha}^{z}\rangle=\frac12\tanh \frac{\beta h_{\alpha}}{2}\,.
\label{GrindEQ_ista2.9}
\end{equation}
The condition of transition to SF-phase is the divergence of  boson Green's
function $\langle\langle S^+ | S^- \rangle\rangle_{q,\omega}$ at zero
frequency and $\vec{q}=0$ (as we approach SF phase boundary from any of
normal phases).

To construct the equations for pseudospin  Green's functions, we use the
linearized equations of  motion for $\vec{\sigma}_{n\alpha}$ operators
\begin{eqnarray}\label{GrindEQ__5_}
&& {\left[\sigma _{l\alpha }^{x} ,\hat{H}\right] = E_{\alpha } \ri\sigma _{\ri\alpha }^{y} -\left\langle \sigma _{\alpha }^{z} \right\rangle
\sum _{n'}\left(J_{ln '}^{\alpha\beta} +J_{n'l}^{\beta\alpha}\right)\ri\sigma _{n'\beta }^{y}  } \,, \nonumber \\
&& {\left[\sigma _{l\alpha }^{y} ,\hat{H}\right] = -E_{\alpha } \ri\sigma _{\ri\alpha }^{x}}  +\left\langle \sigma _{\alpha }^{z} \right\rangle \sum _{n'}\left(J_{ln'}^{\alpha\beta}+J_{n'l}^{\beta\alpha}\right)
\cos \vartheta_{\mathrm{A}} \cos \vartheta_{\mathrm{B}} \ri\sigma_{n'\beta }^{x} \,,\nonumber\\
&& \left[\sigma_{l\alpha }^{z}, \hat{H}\right] = 0
\end{eqnarray}
(these equations were written using RPA decoupling). It is taken into
account that interaction $J_{nn'}^{\alpha\beta}$ (particle hopping) takes
place between lattice sites from different sublattices.

As a result, we obtain the following set of equations for pseudospin Green's
functions
\begin{eqnarray}\label{GrindEQ__7_}
\hbar \omega \langle\langle\sigma _{l\alpha}^{x} |\sigma _{l'\gamma}^{x} \rangle\rangle &=& \ri E_{\alpha} \langle\langle\sigma _{l\alpha}^{y} |\sigma _{l'\gamma}^{x} \rangle\rangle - \ri\left\langle \sigma _{\alpha}^{z} \right\rangle \sum _{n'}\left(J_{ln'}^{\alpha\beta} +J_{n'l}^{\beta\alpha} \right)\langle\langle\sigma _{n'\beta}^{y} |\sigma _{l'\gamma}^{x}
  \rangle\rangle, \nonumber \\
\hbar \omega \langle\langle\sigma _{l\alpha}^{y} |\sigma _{l'\gamma}^{x} \rangle\rangle &=& -\ri\frac{\hbar }{2\pi } \delta _{ll'}\delta_{\alpha\gamma} \left\langle
\sigma _{\alpha}^{z}
\right\rangle -\ri E_{\alpha} \langle\langle\sigma _{l\alpha}^{x} |\sigma _{l'\gamma}^{x} \rangle\rangle %
+ \ri\left\langle \sigma _{\alpha}^{z} \right\rangle \sum _{n'}\left(L_{ln'}^{\alpha\beta} +L_{n'l}^{\beta\alpha} \right)\langle\langle\sigma _{n'\beta}^{x} |\sigma _{l'\gamma}^{x} \rangle\rangle,\qquad
\end{eqnarray}
where \begin{equation}
 L_{ln'}^{\mathrm{AB}} =J_{\ln '}^{\mathrm{AB}} \cos \vartheta _{\mathrm{A}} \cos
\vartheta _{\mathrm{B}}\,. \label{GrindEQ_ista2.12}
\end{equation}
After Fourier transformation of pseudospin interaction matrix
\begin{eqnarray}
J\left(\vec{q}\right) &=& \sum _{n-n'}\left(J_{nn'}^{\mathrm{AB}} +J_{n'n}^{BA} \right){\rm e}^{\ri\vec{q}\left(\vec{R}_{nA} -\vec{R}_{n'B} \right)}
\label{GrindEQ__9_}
\end{eqnarray}
as well as Green's functions $\langle\langle \sigma ^{\alpha } |\sigma ^{\beta } \rangle\rangle $ we
obtain, in particular, the following equations
\begin{eqnarray} \label{GrindEQ__10_}
 \hbar \omega G_{\mathrm{AA}}^{xx} &=& \ri E_{\mathrm{A}} G_{\mathrm{AA}}^{yx} -\ri \left\langle \sigma _{\mathrm{A}}^{z} \right\rangle J\left(\vec{q}\right)G_{\mathrm{BA}}^{yx}\,,  \nonumber \\
 \hbar \omega G_{\mathrm{AA}}^{yx} &=& -\ri \frac{\hbar }{2\pi } \left\langle \sigma _{\mathrm{A}}^{z} \right\rangle -\ri E_{\mathrm{A}} G_{\mathrm{AA}}^{xx} +\ri \left\langle \sigma _{\mathrm{A}}^{z} \right\rangle L\left(\vec{q}\right)G_{\mathrm{BA}}^{xx}\,, \nonumber \\
 \hbar \omega G_{\mathrm{BA}}^{xx} &=& \ri E_{\mathrm{B}} G_{\mathrm{BA}}^{yx} -\ri \left\langle \sigma _{\mathrm{B}}^{z} \right\rangle J\left(\vec{q}\right)G_{\mathrm{AA}}^{yx}\,, \nonumber\\[1ex]
 \hbar \omega G_{\mathrm{BA}}^{yx} &=& -\ri E_{\mathrm{B}} G_{\mathrm{BA}}^{xx} +\ri \left\langle \sigma _{\mathrm{B}}^{z} \right\rangle L\left(\vec{q}\right)G_{\mathrm{AA}}^{xx} \,.
\end{eqnarray}

The system of equations (\ref{GrindEQ__10_}) can be easily solved to obtain the expressions for matrix Green's functions $\langle\langle  \sigma _{\alpha
}^{\mu } |\sigma _{\gamma }^{\nu} \rangle\rangle  _{q,w} $ and $\langle\langle  S_{\alpha }^{\mu }
|S_{\gamma }^{\nu} \rangle\rangle  _{q,w} $, (we can calculate the latter using
relations (\ref{GrindEQ_ista2.4})). Here, $\mu $ and $\nu $ indices denote
$+,-$,$z$ components.

\section{Boson spectrum in normal phase and phase diagrams}

Let us consider the  one-particle boson Green`s function $\langle\langle
b_{\alpha}|b_{\beta}^{+}\rangle\rangle _{q,w}=\langle\langle  S_{\alpha}^{+}|S_{\beta}^{-}\rangle\rangle _{q,w}$.
 In the normal phase case $\langle\langle
S_{\alpha }^{+} |S_{\beta }^{-} \rangle\rangle  _{q,w} =\langle\langle  \sigma _{\alpha }^{+} |\sigma
_{\beta }^{-} \rangle\rangle  _{q,w} $. For $\alpha =\beta$, we have the following result
\begin{equation} \label{GrindEQ__11_}
G_{\alpha \alpha }^{+-} (\vec{q},w)\equiv \langle\langle  \sigma _{\alpha }^{+} |\sigma _{\alpha }^{-} \rangle\rangle  _{q,w} =\frac{\hbar }{\pi } \left\langle \sigma _{\alpha }^{z} \right\rangle
\frac{\hbar \omega -E_{\beta } }{(\hbar \omega -E_{\alpha } )(\hbar \omega -E_{\beta } )-\Phi _{q} }\, ,
\end{equation}
derived from equations (\ref{GrindEQ__10_}). Here, $\Phi _{q} =\left\langle \sigma _{\mathrm{A}}^{z} \right\rangle \left\langle
\sigma _{\mathrm{B}}^{z} \right\rangle J^{2} \left(\vec{q}\right)$.

The boson excitation spectrum is defined from the poles of the
$G_{\alpha\alpha}^{+-}$ function
\begin{equation} \label{GrindEQ__13_}
\varepsilon _{1,2}^{(NO)} (\vec{q})=h\pm \sqrt{\delta ^{2} +\left\langle \sigma _{\mathrm{A}}^{z} \right\rangle \left\langle \sigma _{\mathrm{B}}^{z} \right\rangle J^{2} (\vec{q})}\,.
\end{equation}
We have introduced the general notations $h=\frac{E_{\mathrm{A}} +E_{\mathrm{B}} }{2} $; $\delta
=\frac{E_{\mathrm{A}} -E_{\mathrm{B}} }{2}$. In normal phases $h=\frac{h_{\mathrm{A}}+h_{\mathrm{B}}}{2}$;
$\delta=\frac{h_{\mathrm{A}}-h_{\mathrm{B}}}{2}$.

\pagebreak
The features of the obtained spectrum may vary depending on the values of  the
model parameters:
\paragraph{For $\delta =0$} (A and B positions are equivalent; crystal is not
split to sublattices and the unit cell is two times smaller):

\begin{equation}
\varepsilon _{1,2}^{(NO)} (\vec{q})=h\pm \left|\left\langle \sigma ^{z} \right\rangle \right|J(\vec{q}), \qquad \left\langle \sigma ^{z} \right\rangle =\frac{1}{2} \tanh \frac{\beta h}{2}\,.
\end{equation}
There is only one band $\varepsilon (\vec{q})=h-\left\langle \sigma ^{z}
\right\rangle J(\vec{q})$ inside the two times bigger Brillouin zone.

\paragraph{For $\delta \ne 0;$ $\delta >0.$}

There are two bands in this case. The edges of the bands are defined by the
inequalities which depend on the sign of $\left\langle \sigma _{\mathrm{A}}^{z}
\right\rangle \left\langle \sigma _{\mathrm{B}}^{z} \right\rangle =\frac{1}{4} _{}
\tanh _{} \frac{\beta}{2}(h+\delta ) _{} \tanh _{} \frac{\beta}{2}
(h-\delta )$ expression:

$\begin{array}{l} {h+\delta <\varepsilon _{1} (\vec{q})<h+\sqrt{\delta ^{2}
+\left\langle \sigma _{\mathrm{A}}^{z} \right\rangle \left\langle \sigma _{\mathrm{B}}^{z}
\right\rangle J^{2} (0)} } \\ {h-\sqrt{\delta ^{2} +\left\langle \sigma
_{\mathrm{A}}^{z} \right\rangle \left\langle \sigma _{\mathrm{B}}^{z} \right\rangle J^{2} (0)}
<\varepsilon _{2} (\vec{q})<h-\delta } \end{array}$  $\left. \begin{array}{c} {} \\
{} \\ {} \\ {} \end{array}\right\}$ $ $$\left\langle \sigma _{\mathrm{A}}^{z}
\right\rangle \left\langle \sigma _{\mathrm{B}}^{z} \right\rangle >0$

and

$\begin{array}{l} {h+\sqrt{\delta ^{2} +\left\langle \sigma _{\mathrm{A}}^{z}
\right\rangle \left\langle \sigma _{\mathrm{B}}^{z} \right\rangle J^{2} (0)} _{}
<\varepsilon _{1} (\vec{q})<h+\delta } \\ {h-\delta <\varepsilon _{2}
(\vec{q})<_{} h-\sqrt{\delta ^{2} +\left\langle \sigma _{\mathrm{A}}^{z} \right\rangle
\left\langle \sigma _{\mathrm{B}}^{z} \right\rangle J^{2} (0)} } \end{array}$  $\left.
\begin{array}{c} {} \\ {} \\ {} \\ {} \end{array}\right\}$ $ $$\left\langle
\sigma _{\mathrm{A}}^{z} \right\rangle \left\langle \sigma _{\mathrm{B}}^{z} \right\rangle <0$.

\begin{figure}[!h]
\centerline{
\includegraphics[width=0.48\columnwidth]{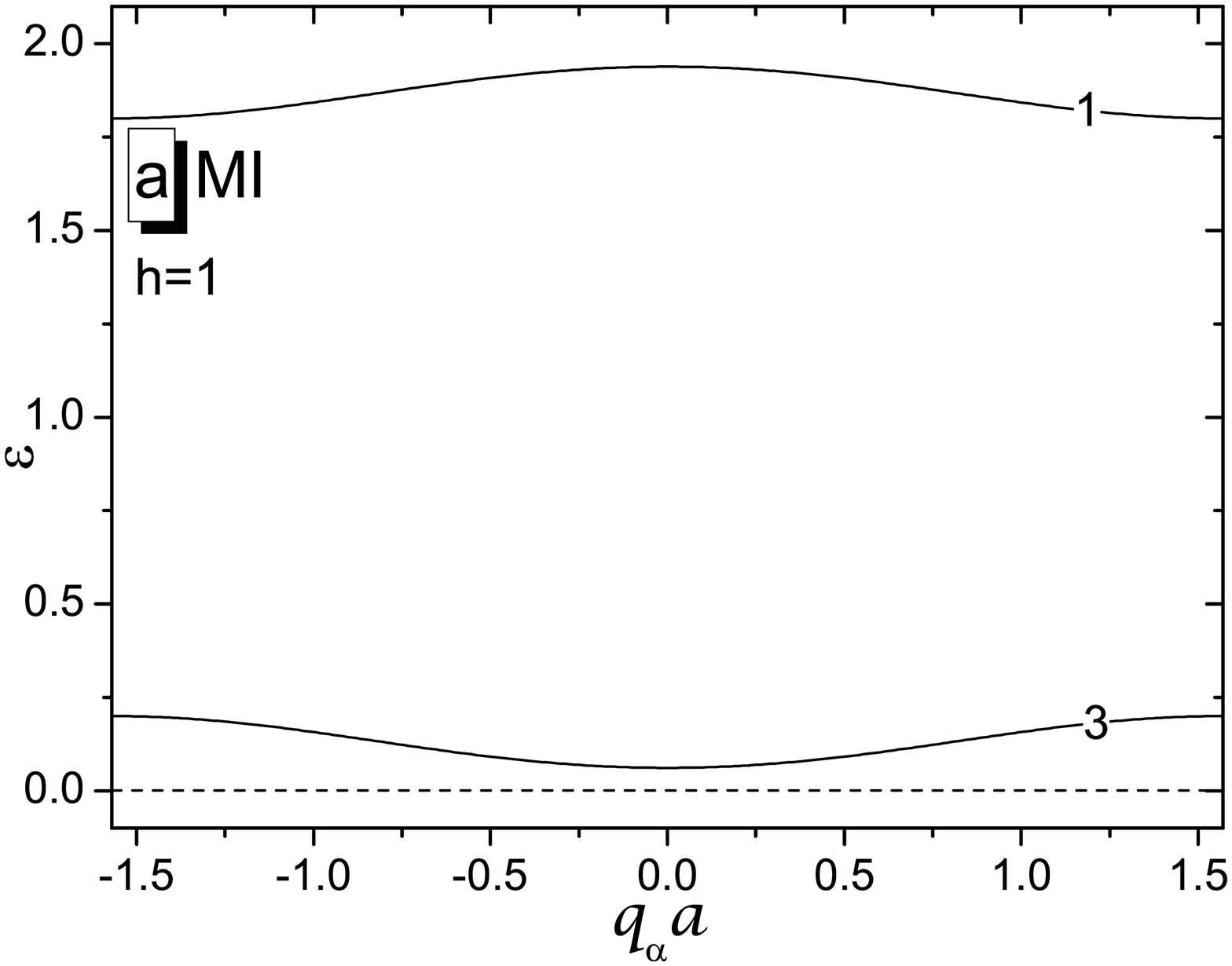}
\includegraphics[width=0.485\columnwidth]{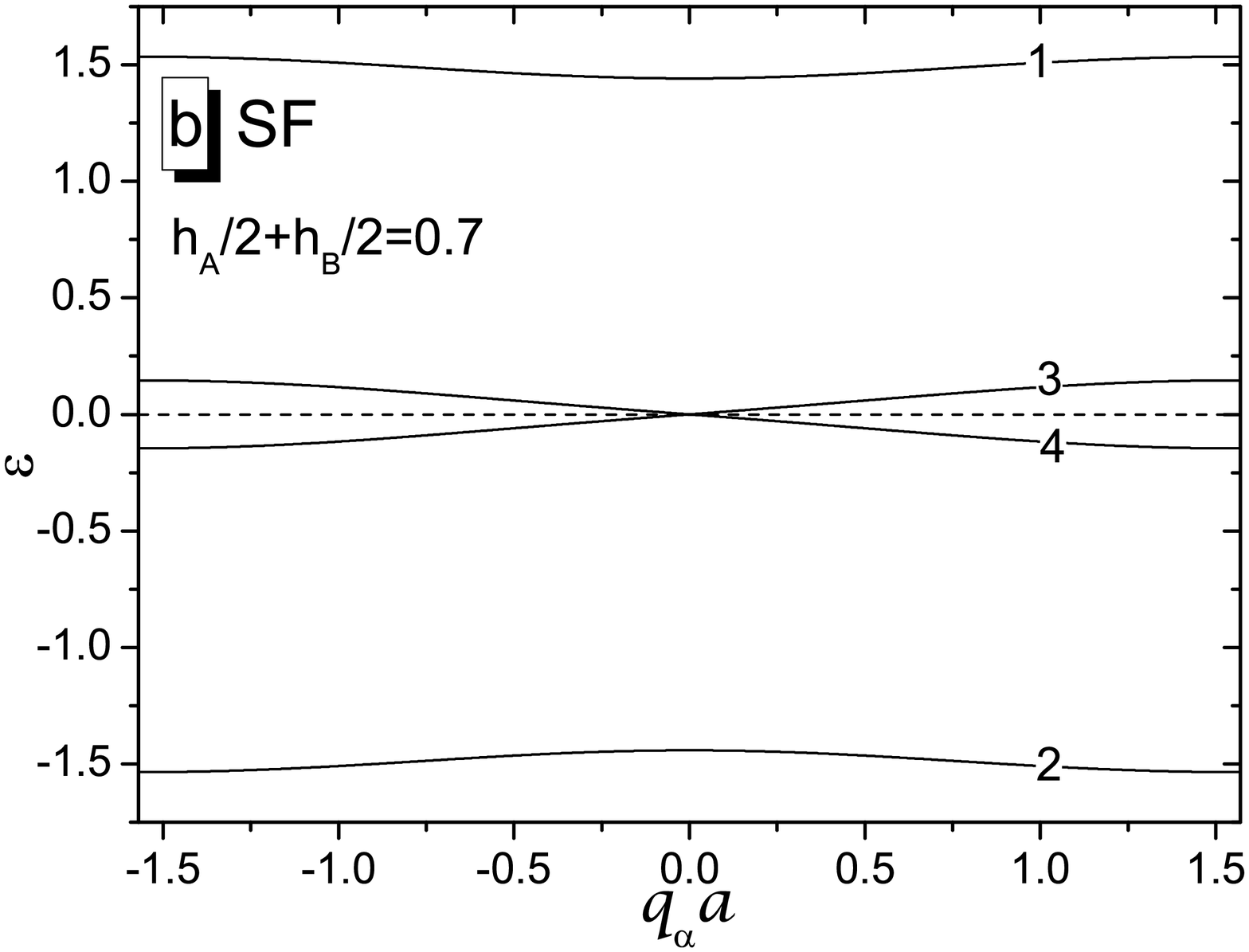}
}
\centerline{
\includegraphics[width=0.48\columnwidth]{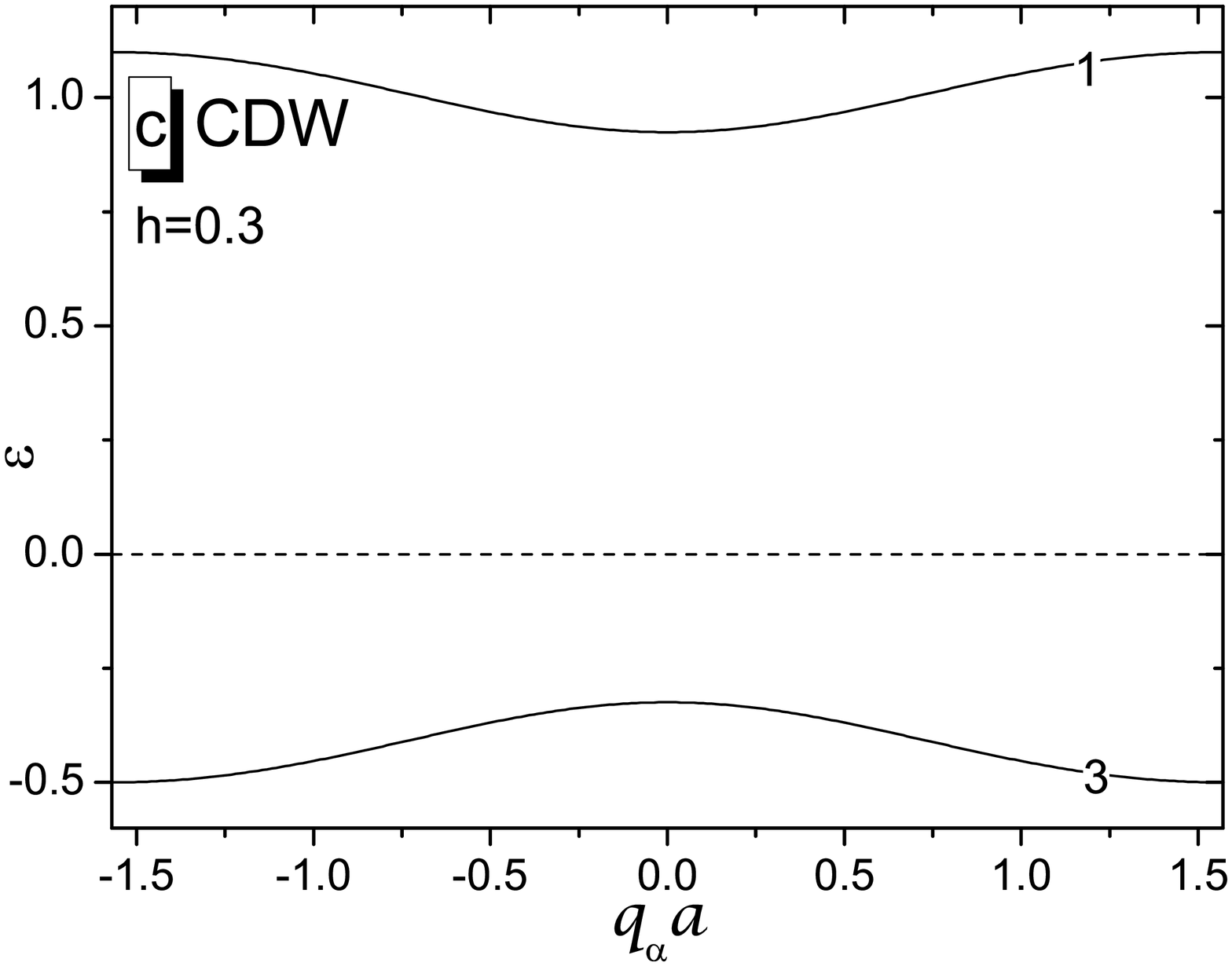}
\includegraphics[width=0.48\columnwidth]{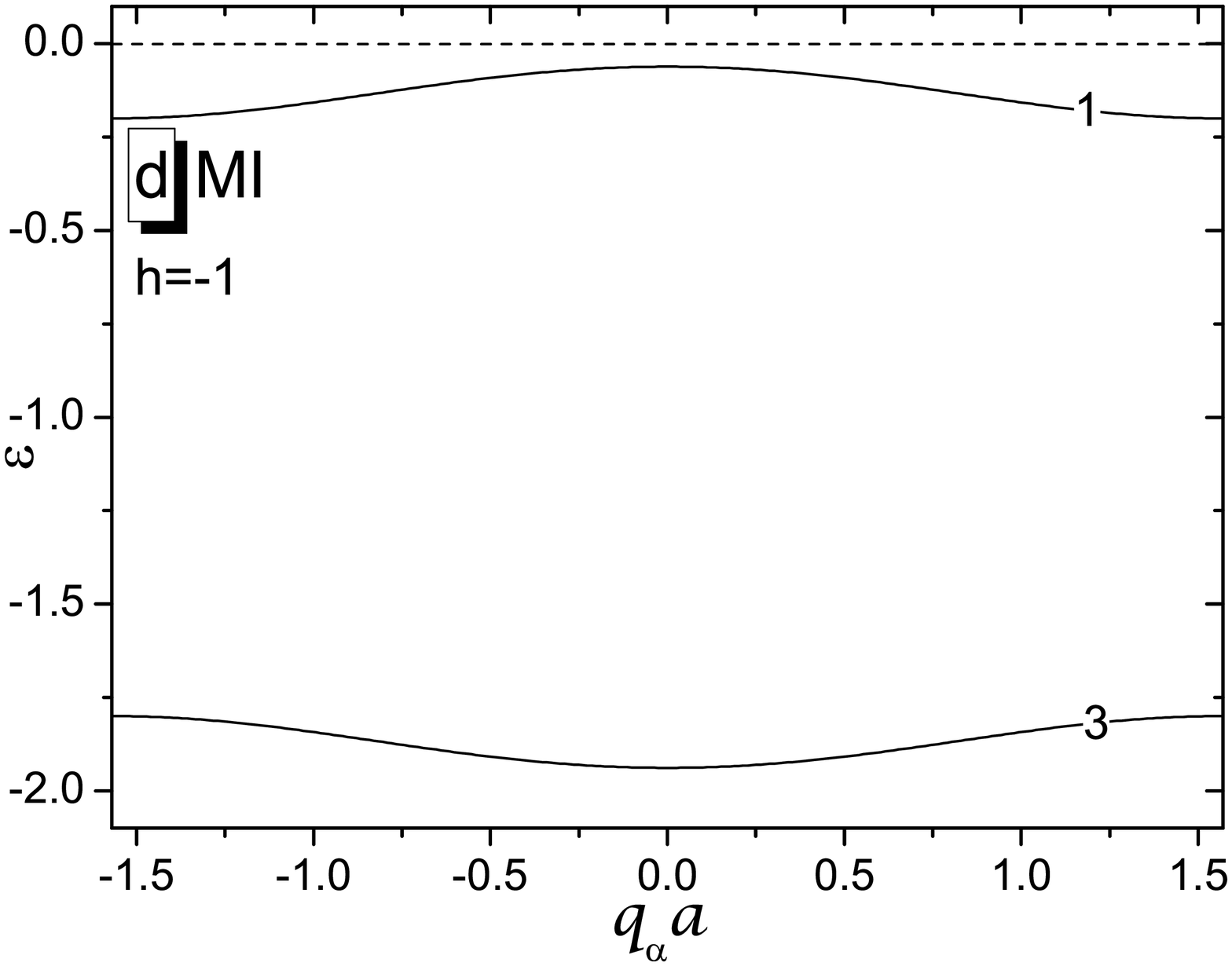}
}
\caption{Dispersion laws $\varepsilon(q)$ for different phases. Dashed line denotes the chemical potential level. The Fourier transform $J(\vec{q})=\frac{1}{z}J(0)\sum\limits_{\alpha=1}^{z}\cos q_{\alpha}a$ is used with the aim of  illustration. $J(0)$ is chosen as the energy unit. $T \equiv 1/\beta =0.05, \delta=0.8$. The numbers indicate the corresponding branches. For SF phase: 1 -- $\varepsilon_1$, 2 -- $\varepsilon_2$, 3 -- $\varepsilon_3$, 4 -- $\varepsilon_4$. For MI and CDW phases: 1 -- $\varepsilon_1$, 3 -- $\varepsilon_2$.} \label{spectra}
\end{figure}
In the first case ($\langle \sigma _{\mathrm{A}}^{z} \rangle \langle
\sigma _{\mathrm{B}}^{z} \rangle >0$), which holds for $h-\delta >0$, two
different bands always exist; the gap between these bands disappears as
$\delta \to 0$. The chemical potential (which is located on the energy scale
at $\varepsilon =0$ point) is placed either higher or lower than the bands
$\varepsilon _{1} (\vec{q})$ and $\varepsilon _{2} (\vec{q})$ [figures~\ref{spectra}~(a), \ref{spectra}~(d)]. In the second case ($ \langle \sigma
_{\mathrm{A}}^{z}  \rangle  \langle \sigma _{\mathrm{B}}^{z}  \rangle <0$), which
corresponds to the inequalities $h-\delta <0$; $h+\delta >0$, two different
bands exist only at $\delta
>\sqrt{ | \langle \sigma _{\mathrm{A}}^{z}  \rangle  \langle \sigma
_{\mathrm{B}}^{z}  \rangle  |}J(0)$. The gap disappears when this condition
is violated [at $T=0$ this happens at $\delta =\delta _{\mathrm{c}} \equiv \frac{1}{2}
J(0)$]. When the bands are separated in normal phase, the chemical potential is
located between the bands [figure~\ref{spectra}~(c)]. The instability connected
with SF transition takes place when the level of chemical potential touches
the edge of one of the bands  that may be driven either by the temperature,
chemical potential or energy difference $\delta$ change. At $J(0)>0$ ($t_{ij}
>0$), this always happens in the $\vec{q}=0$ point. The condition for this is
as follows:
\begin{equation} \label{GrindEQ__16_}
h^{2} =\delta ^{2} +\left\langle \sigma _{\mathrm{A}}^{z} \right\rangle \left\langle \sigma _{\mathrm{B}}^{z} \right\rangle J^{2} (0).
\end{equation}

\begin{figure}[ht]
\includegraphics[width=0.455\textwidth]{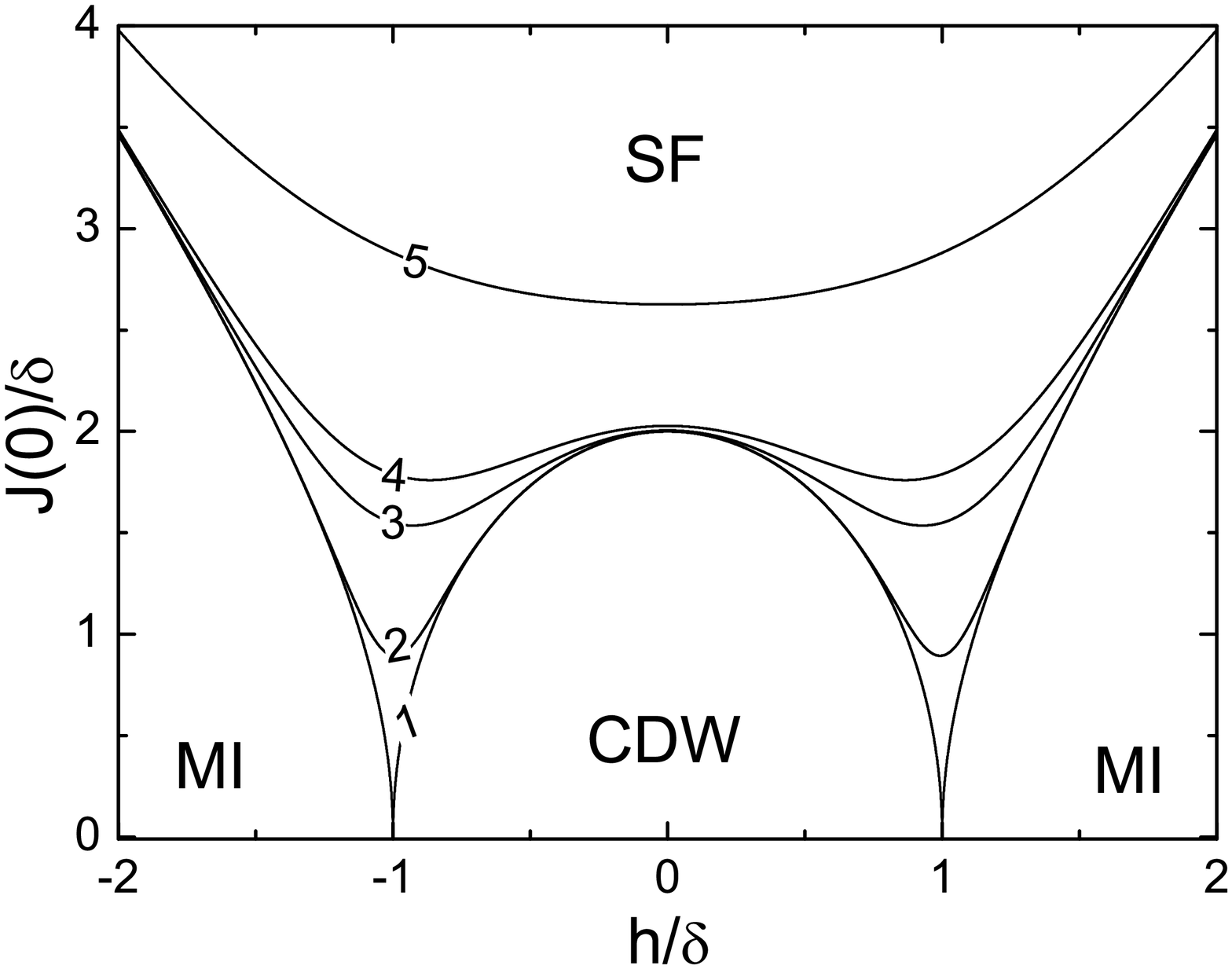}%
\hfill%
\includegraphics[width=0.48\textwidth]{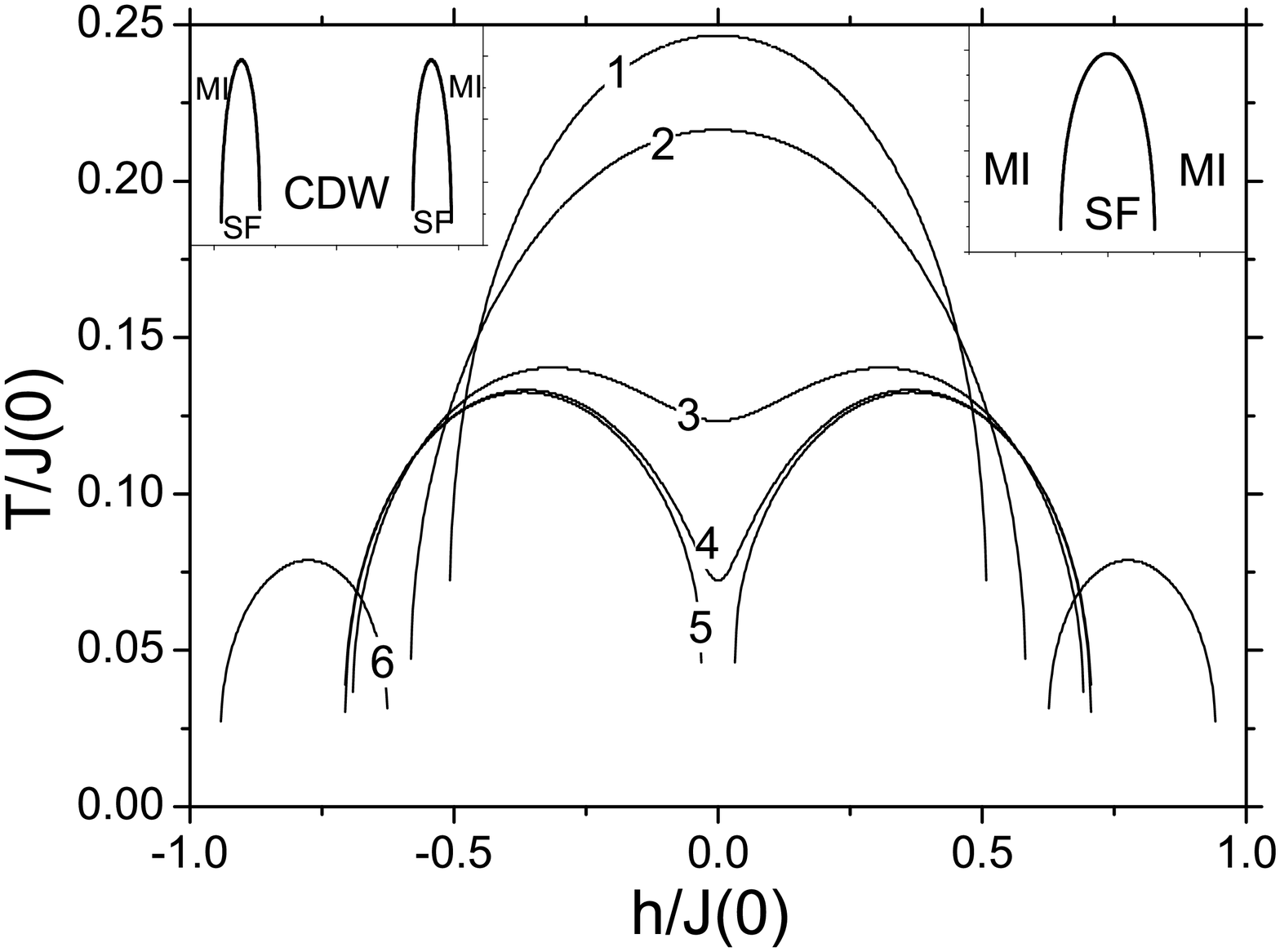}%
\\%
\parbox[t]{0.48\textwidth}{%
\caption{Phase diagram of two-sublattice model of hard-core bosons for different temperatures: 1.~$T=0.00005$, 2.~$T=0.05$, 3.~$T=0.15$, 4.~$T=0.2$, 5.~$T=0.5$. Energy quantities are measured in units of $\delta =  (\varepsilon _{\mathrm{A}} -\varepsilon _{\mathrm{B}} )/{2}$. \label{diag1}}%
}
\hfill%
\parbox[t]{0.48\textwidth}{%
\caption{Phase diagram $(T, h)$ at various values of $\delta$: 1.~$\delta=0.1$, 2.~$\delta=0.3$, 3.~$\delta=0.48$, 4.~$\delta=0.499$, 5.~$\delta=0.501$, 6.~$\delta=0.8$. Energy quantities are measured in units of $J(0)$.\label{diag2}}%
}%
\end{figure}

Two equations derived from this relation allow us to construct the phase
diagrams in $\left(J(0),h\right)$ and $\left(T,h\right)$ planes that show the
areas of SF and normal (MI, CDW) phases. Diagram in figure~\ref{diag1}
illustrates the change of the shape of phase boundary curve on
$\left(J(0),h\right)$ plane as the temperature increases (at $T=0$, the phase
boundary curve corresponds to the one obtained in \cite{iskin,hen2}). The definitive boundary between MI and CDW regions exists only at zero temperature. In this case, MI and CDW states can be interpreted as different phases. When one departs from $T=0$ limit, this boundary disappears and one may observe a single normal phase. However, this normal phase is close to either MI or CDW phases in different regions of phase diagram (also see below).

If the existing critical value of the difference of sublattice local energies $\left(\delta =\delta _{\mathrm{c}} \right)$ is exceeded, it leads to the splitting of the SF-phase area on
$\left(T,h\right)$ plain (figure~\ref{diag2}). This result is in agreement with
the papers mentioned above, where all calculations were performed only at
$T=0$. Therefore, at $\delta >\delta _{\mathrm{c}}$, there are two critical points for
$T\ne 0$.

 For intermediate values of chemical potential, the normal phase is similar to the charge ordered phase (CDW) while at large positive (or negative) values of $h$ this phase is of Mott-insulator (MI) type.
 This conclusion is confirmed by one-particle spectral density $\rho _{\alpha } (\omega )$
 calculations. We use the relation
\begin{equation} \label{GrindEQ__17_}
 \rho _{\alpha } (\omega )=-\frac{1}{N} \sum _{q}2\Im\langle\langle  S_{\alpha }^{+} |S_{\alpha }^{-} \rangle\rangle  _{q,\omega +\ri\varepsilon }
 =\frac{2}{N} \sum _{q}\left\langle \sigma _{\alpha }^{z} \right\rangle \left\{A_{1}^{\alpha } (\vec{q})\delta \left[\omega -\frac{\varepsilon _{1} (\vec{q})}{\hbar } \right]+A_{2}^{\alpha }
 (\vec{q})\delta \left[\omega -\frac{\varepsilon _{2} (\vec{q})}{\hbar } \right]\right\}, \nonumber
\end{equation}
which follows from the decomposition into partial fractions.

Here,
\[
A_{1,2}^{\mathrm{A}} (\vec{q})=\frac{1}{2} \pm \frac{\delta }{2\sqrt{\delta ^{2} +\Phi_{q}} } \,,
\]
while expression for $A_{1,2}^{\mathrm{B}} (\vec{q})$ is derived from $A_{1,2}^{\mathrm{A}}
(\vec{q})$ by  $A\rightleftarrows B$ $\left(\delta \to -\delta \right)$
substitution.

Using non-perturbative density of states
\[
\rho _{0} (z)=\frac{1}{N} \sum _{q}\delta \left[z-J(\vec{q})\right],
\]
 we can rewrite the expression (\ref{GrindEQ__17_}) for $\alpha =A$
\begin{eqnarray} \label{GrindEQ__20_}
\rho _{\mathrm{A}} (\omega ) &=& 2 \langle \sigma^z_a \rangle \int _{-J(0)}^{J(0)}\rd z\rho _{0} (z)
 \left\{\left(\frac{1}{2} \!+\! \frac{\delta }{2\sqrt{\delta ^{2} \!+\! \left\langle \sigma _{\mathrm{A}}^{z} \right\rangle \left\langle \sigma _{\mathrm{B}}^{z} \right\rangle z^{2} } } \right)\delta \left[\omega \!-\! \frac{1}{\hbar } \left( h \!+\! \sqrt{\delta ^{2} \!+\! \left\langle \sigma _{\mathrm{A}}^{z} \right\rangle \left\langle \sigma _{\mathrm{B}}^{z} \right\rangle z^{2} } \right) \right]\right. \nonumber \\
&&{}+ \left. \left(\frac{1}{2} \!-\! \frac{\delta }{2\sqrt{\delta ^{2} \!+\! \left\langle \sigma _{\mathrm{A}}^{z} \right\rangle \left\langle \sigma _{\mathrm{B}}^{z} \right\rangle z^{2} } } \right)\delta \left[\omega \!-\! \frac{1}{\hbar } \left(h-\sqrt{\delta ^{2} \!+\! \left\langle \sigma _{\mathrm{A}}^{z} \right\rangle \left\langle \sigma _{\mathrm{B}}^{z} \right\rangle z^{2} } \right)\right]\right\}.
\end{eqnarray}
When performing numerical calculations, we use the semi-elliptical function $\rho _{0} (z)=\frac{1}{\pi J^{2} (0)} \sqrt{J^{2} (0)-z^{2} } $.

\begin{figure}[!h]
\centerline{
\includegraphics[width=0.48\columnwidth]{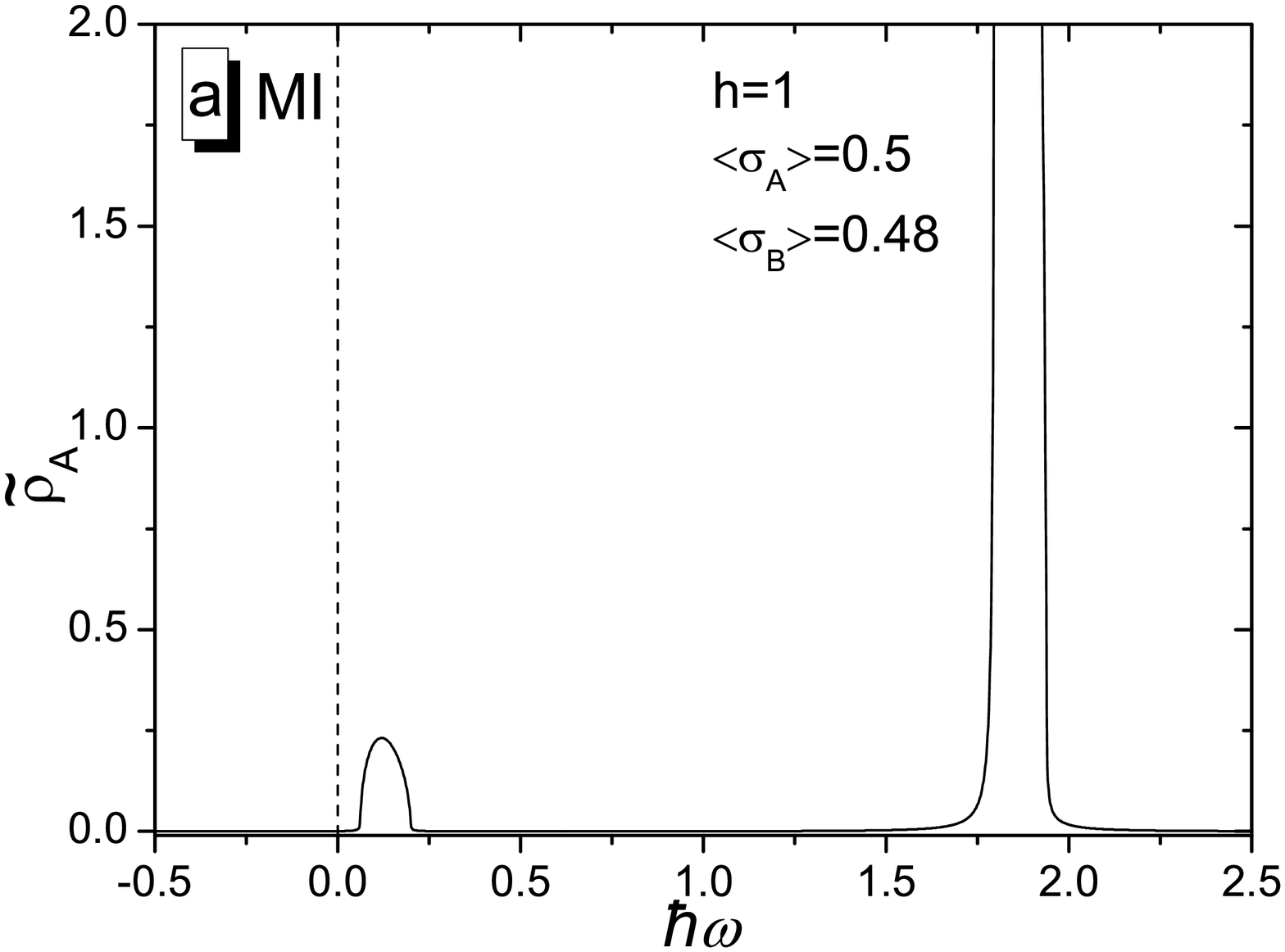}
\includegraphics[width=0.48\columnwidth]{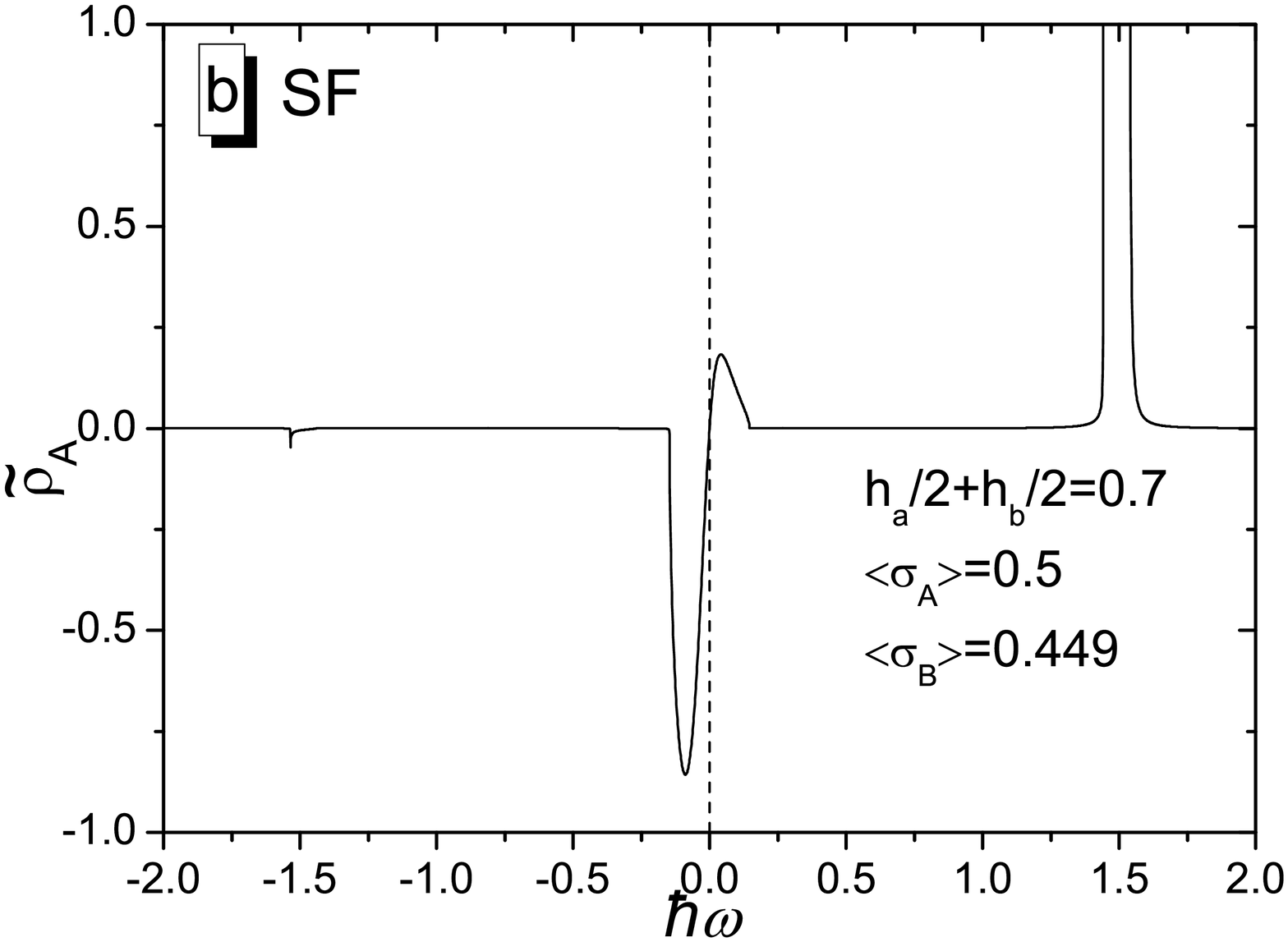}
}
\centerline{
\includegraphics[width=0.48\columnwidth]{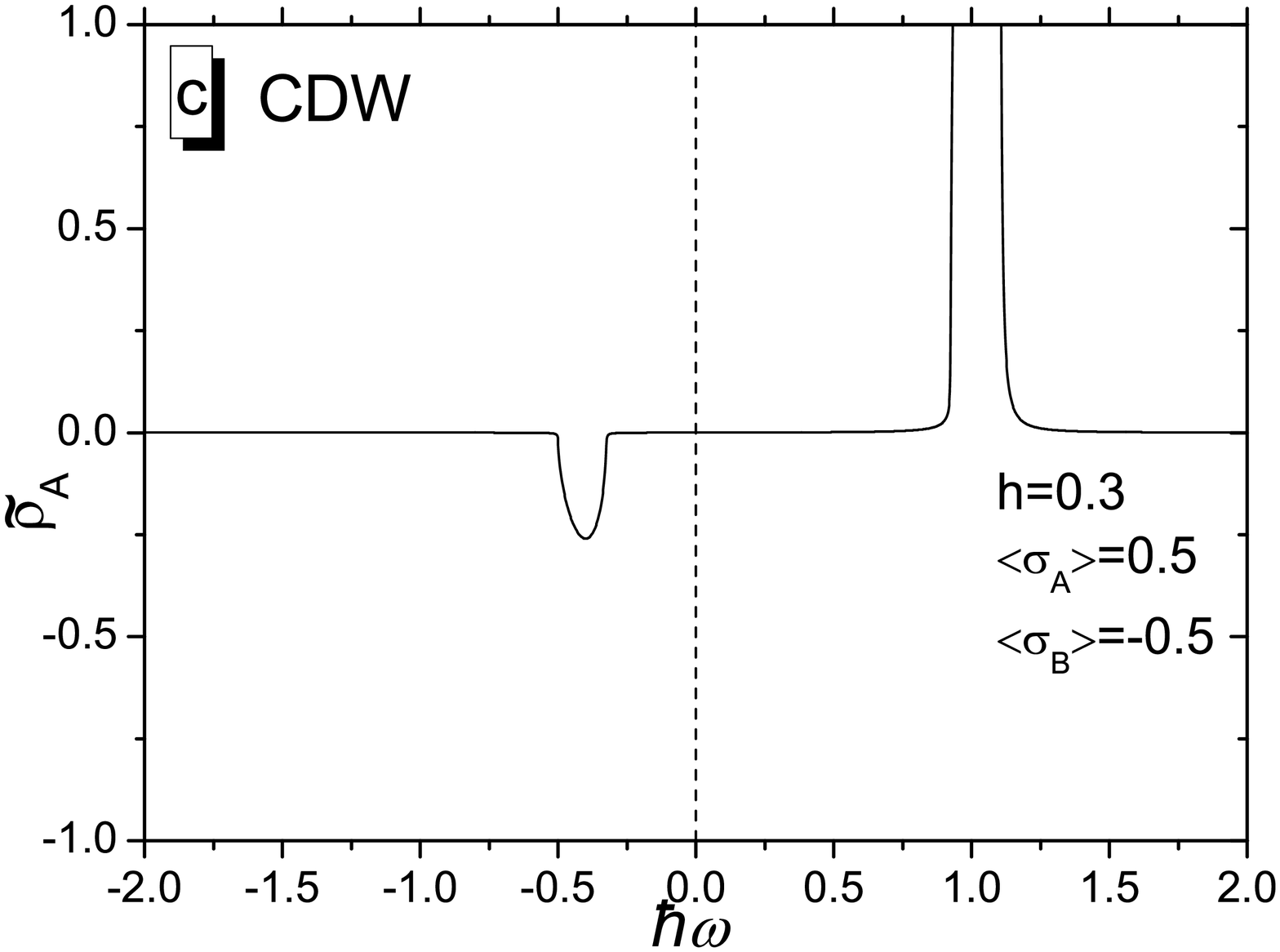}
\includegraphics[width=0.48\columnwidth]{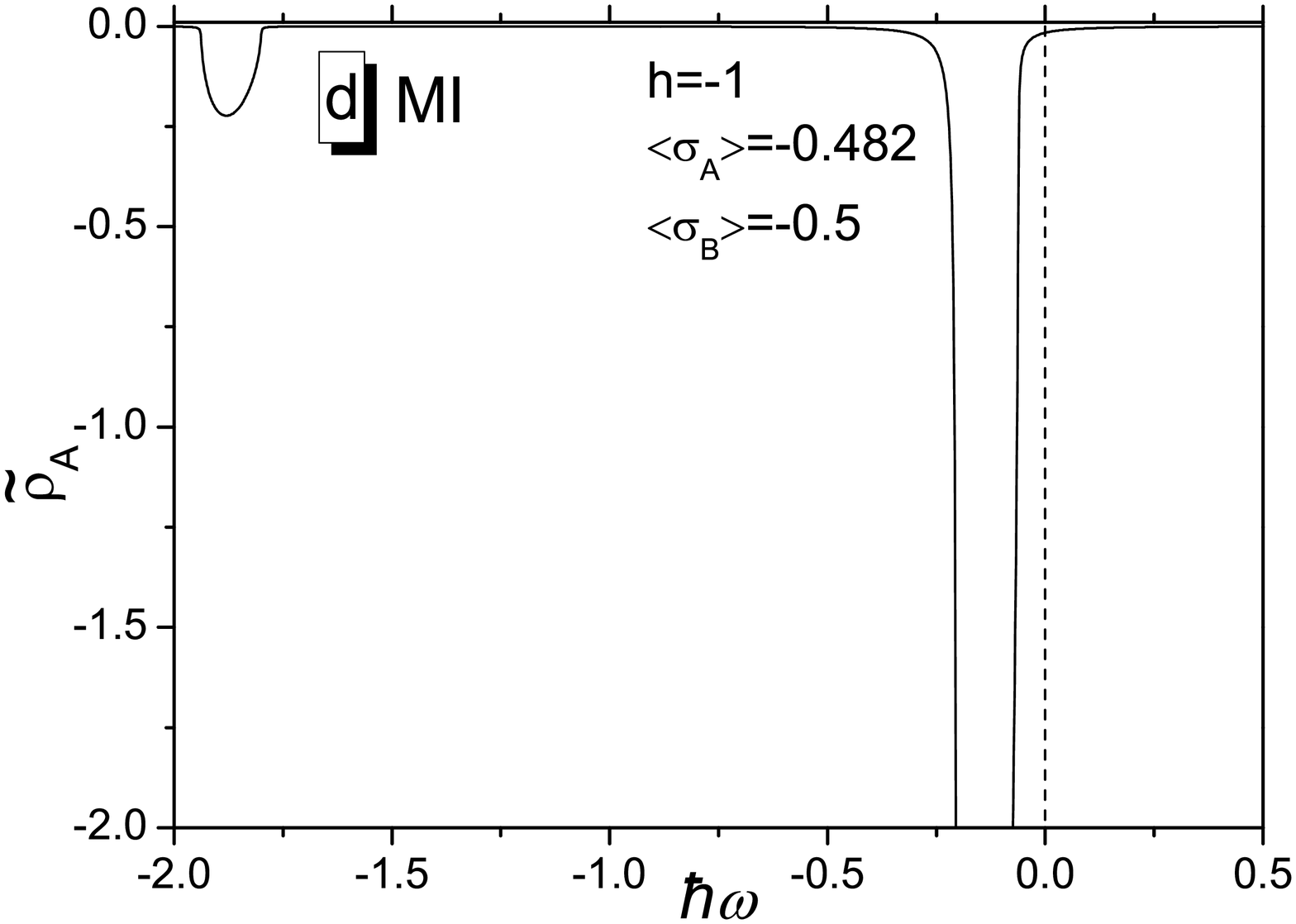}
}
\caption{Spectral density of A-sublattice for different phases. $T=0.05$, $\delta=0.8$. All energy quantities are measured in units of $J(0)$. $\tilde{\rho}_{\mathrm{A}}=\rho_{\mathrm{A}} / \hbar$ is the spectral density as function of energy $\hbar \omega$.}\label{dos}
\end{figure}
Figure~\ref{dos} illustrates spectral density for all phases. For CDW region of normal phase [figure~\ref{dos}~(c)], the chemical potential is located within the
gap between the bands $\rho _{\alpha } (\omega )$; the sign of $\rho _{\alpha
} (\omega )$ function is different in each band [$\rho _{\alpha } (\omega
)<0$ at $\hbar \omega <\mu $ and $\rho _{\alpha } (\omega )>0$ at $\hbar
\omega >\mu $]. For MI region [figures~\ref{dos}~(a), \ref{dos}~(d)] the chemical
potential is at the same side of both bands $\rho _{\alpha } (\omega )$.

The values of $\sigma_{\mathrm{A}}^z$ and $\sigma_{\mathrm{B}}^z$ averages presented in figures~\ref{dos}~(a)--\ref{dos}~(d) are very close to those at $T=0$. We observe the modulated occupancy $n_{\mathrm{A}}=\frac{1}{2}-\sigma_{\mathrm{A}}^z=0, n_{\mathrm{B}}=\frac{1}{2}-\sigma_{\mathrm{B}}^z=1$ in CDW-like case [figure~\ref{dos}~(c)]. Contrary to this, in MI-like cases, this occupancy is either close to zero or unity depending on the chemical potential value. The latter two possibilities are illustrated in figure~\ref{dos}~(a) ($n_{\mathrm{A}} \approx n_{\mathrm{B}} \approx 0$ when $\mu$ is positioned below the energy bands) and figure~\ref{dos}~(d) ($n_{\mathrm{A}} \approx n_{\mathrm{B}} \approx 1$ when $\mu$ is placed above the bands).

\section{Excitation spectrum in SF phase}

In the case of a phase with BE-condensate (SF phase), when
$\sin\vartheta_{\mathrm{A}}\neq0$, $\sin\vartheta_{\mathrm{B}}\neq0$,
\begin{equation} \label{GrindEQ_ista4.1}
\langle\langle  S_{\mathrm{A}}^{+}|S_{\mathrm{A}}^{-}\rangle\rangle _{q,w}=\frac{\hbar}{2\pi}\langle\sigma_{\mathrm{A}}^{z}\rangle
\frac{P_{q}^{\mathrm{A}}}{(\hbar^{2}\omega^{2}-E_{\mathrm{A}}^{2})(\hbar^{2}\omega^{2}-E_{\mathrm{B}}^{2})-2M_{q}\hbar^{2}\omega^{2}-2N_{q}E_{\mathrm{A}}E_{\mathrm{B}}+M_{q}^{2}}\,,
\end{equation}
where
\begin{equation} \label{GrindEQ_ista4.2}
P_{q}^{\mathrm{A}}(\hbar\omega)=\left[E_{\mathrm{A}}\left(\cos^{2}\vartheta_{\mathrm{A}}+1\right)+2\hbar\omega\cos\vartheta_{\mathrm{A}}\right]\left(\hbar^{2}\omega^{2}-E_{\mathrm{A}}^{2}\right)-
2\hbar\omega M_{q}\cos\vartheta_{\mathrm{A}}+\tilde{\Phi}_{q}^{\mathrm{A}}E_{\mathrm{B}}\,,
\end{equation}
and the  following notations are introduced:
\begin{eqnarray} \label{GrindEQ_ista4.3}
M_{q}=\Phi_{q}\cos\vartheta_{\mathrm{A}}\cos\vartheta_{\mathrm{B}}\,,\qquad N_{q}=\frac12\Phi_{q}\left(1+\cos^{2}\vartheta_{\mathrm{A}}\cos^{2}\vartheta_{\mathrm{B}}\right), \qquad
\tilde{\Phi}_{q}^{\mathrm{A}}= \Phi_{q}\cos^{2}\vartheta_{\mathrm{A}}\left(1+\cos^{2}\vartheta_{\mathrm{B}}\right)
\end{eqnarray}
(the replacement $A\rightleftarrows B$ gives an expression for  the  $\langle\langle
S_{\mathrm{B}}^{+}|S_{\mathrm{B}}^{-}\rangle\rangle _{q,w}$ function).

The boson spectrum consists now of four branches
\begin{equation} \label{GrindEQ_ista4.4}
\varepsilon_{1,2}^{\mathrm{(SF)}}(\vec{q})=\pm\left(P_{q}+Q_{q}\right)^{1/2},\qquad
\varepsilon_{3,4}^{\mathrm{(SF)}}(\vec{q})=\pm\left(P_{q}-Q_{q}\right)^{1/2}.
\end{equation}
Here,
\begin{eqnarray} \label{GrindEQ_ista4.5}
P_{q}=\frac12\left(E_{\mathrm{A}}^{2}+E_{\mathrm{B}}^{2}\right)+M_{q}\,, \qquad
 Q_{q}=\left[\frac{1}{4} \left(E_{\mathrm{A}}^{2}-E_{\mathrm{B}}^{2}\right)^{2}+2N_{q}E_{\mathrm{A}}E_{\mathrm{B}}+M_{q}\left(E_{\mathrm{A}}^{2}-E_{\mathrm{B}}^{2}\right)\right]^{1/2}.
\end{eqnarray}
Energies $E_{\mathrm{A}}$ and $E_{\mathrm{B}}$, as well as averages
$\langle\sigma_{\mathrm{A}}^{z}\rangle$ and $\langle\sigma_{\mathrm{B}}^{z}\rangle$ are
determined now as solutions of equations (\ref{GrindEQ_ista2.7}) and
(\ref{GrindEQ__9_}). Regions  of existence of SF phase are shown in
figures~\ref{diag1}, \ref{diag2}. The dispersion curves
$\varepsilon_{1..4}^{\mathrm{(SF)}}(\vec{q})$ are present in figure~{\ref{spectra}~(b)} for
certain values of $h$ and $\delta$ parameters.

The presence of  branches with linear dispersion at small values of $q$
[$\varepsilon_{3}(\vec{q})$ and $\varepsilon_{4}(\vec{q})$ in the case
presented in figure~{\ref{spectra}~(b)}] is the specific feature of SF phase;
their energy goes to zero in the point of the location of chemical potential.
This peculiarity of spectrum is well known from investigations of the
simple hard-core boson model \cite{micnas}. However, in our case, at
$\varepsilon_{\mathrm{A}}\neq\varepsilon_{\mathrm{B}}$, the additional gapped branch
[$\varepsilon_{2}(\vec{q})$ in figure~{\ref{spectra}~(b)}] appears in the negative
energy region.

Similarly to the  normal phase case, one can  perform calculations of the
boson spectral density $\rho_{\alpha}(\omega)$. Using decomposition of
expression (\ref{GrindEQ_ista4.1}) into partial fractions, we obtain
\begin{equation} \label{GrindEQ_ista4.6}
\rho_{\alpha}(\omega)=\frac{2}{N}\sum_{q}\langle \sigma_{\alpha}^{z}\rangle\sum_{i=1}^{4}A_{i}^{\alpha}(\vec{q}) \delta\left(\omega
-\frac{\varepsilon_{i}(\vec{q})}{\hbar}\right),
\end{equation}
where
\begin{equation} \label{GrindEQ_ista4.7}
A_{i}^{\alpha}(\vec{q})=\frac{P_{q}^{\alpha}\left(\hbar\omega= \varepsilon_{i}(\vec{q})\right)}{4Q_{q}\varepsilon_{i}(\vec{q})}\,.
\end{equation}
It is easy to obtain an expression like (\ref{GrindEQ__20_}) passing to
integration with the $\rho_{0}(z)$ density of states. The contributions
from  all four bands are present in the  total spectral density.

The  plots  of the $\rho_{\mathrm{A}}(\omega)$ functions in the case of SF phase are
presented in figure~{\ref{dos}~(b)}. For branches with linear dispersion
$[\varepsilon_{3,4}(\vec{q})]$, the  spectral density changes its sign in the
point $\hbar\omega=0$ (at that point the chemical potential is located). The
change of the spectral density shape at MI $\rightarrow$ SF transition, when
we observe the appearance of the negative branch of $\rho _{\alpha } (\omega
)$ [figure~\ref{dos}~(b)], corresponds to the results obtained in
\cite{ohashi,menotti} as well as to the ones obtained for generalized
hard-core boson model with excited states transfer \cite{velychko}.
Additional branch $\varepsilon_{2}(\vec{q})$ that appears in SF phase is
characterized by a negative spectral density. Its intensity (at the chosen
values of $h$ and $\delta$ parameters) is small. Qualitatively, this shape
of the  $\rho_{\alpha}(\omega)$ function is specific for the
Bose-Hubbard model \cite{Dupuis}. However, contrary to the  standard case,
where additional branches separated by gaps exist due to the  local energy
splitting (caused by the Hubbard repulsion of  bosons),  in our
two-sublattice model such an  effect is a consequence of the energy
non-equivalence of sublattices.

The behaviour of $\rho _{\mathrm{A}}(\omega)$ function is in agreement with the
results of numerical calculations performed in \cite{vorobyov} with exact
diagonalization technique for one-dimensional $\left(d=1\right)$ chain
structures. In \cite{vorobyov}, the authors take into account the two-particle
interaction between nearest neighbouring sites. This interaction forms the
effective internal field which is similar to the field $\delta $ considered
here, and both fields are responsible for the appearance of CDW-like phase.
The shape of spectral densities in various phases, obtained here, lets one
identify the equilibrium states on phase diagrams (diagrams of state)
obtained numerically for $d=1$.

\section{Conclusions}

 Within the random phase approximation, we have calculated the spectral densities of a  two-sublattice model of hard-core bosons and analyzed
  the features of the boson single-particle spectrum in various phases. These features
  are connected with the position of the chemical potential level. It is placed:

\begin{itemize}
  \item  within the gap between two boson bands in the case when normal phase is similar to the       charge-ordered  (CDW) phase;

  \item  above (or below) both bands in the case when normal phase is similar to the Mott insulator (MI) phase;

  \item  within a certain  boson band,  for SF phase (the  phase with
      BE condensate); the  additional boson bands appear in this case.
\end{itemize}

We have obtained the equation that describes the transition to the SF phase
 and have built the corresponding phase diagrams
at various temperatures and at different  values of  energy difference
$\delta=\frac12(\varepsilon_{\mathrm{A}}-\varepsilon_{\mathrm{B}})$. The temperature increase
leads to the gradual vanishing of the difference between CDW-like and MI-like
modifications of normal phase; there are no border lines separating them. SF-phase region also
decreases with the temperature increase;at the  same time, two regions of the
SF phase, which exist at $T=0$ and at a fixed value of $\delta$, join together. On
the  other hand, a similar effect takes place for fixed temperature at the
decrease of $\delta$. At high values of $\delta$, there are two critical
points in  which the  SF phase disappears at an increase of temperature.
When $\delta$ decreases, only one central critical point remains.

At the same time, it should be mentioned that nonzero value of $\delta$ is the main reason for the appearance of the CDW-like state in our system. We have not included direct intersite interactions between particles into consideration. This kind of interaction may induce the phase transition into ``true'' CDW phase.

More elaborate study of the boson spectrum reconstruction at the  transitions
between different regions in  phase diagrams and the  change of their
topology remains an interesting task. It is worthy of special attention.

\ukrainianpart
\title{Енергетичний спектр i фазовi дiаграми двопiдґраткової моделi
жорстких бозонiв}
\author{I.В. Стасюк, О. Воробйов}
\address{Інститут фізики конденсованих систем НАН України,\\
вул. Свєнціцького, 1, 79011 Львів, Україна}
\makeukrtitle

\begin{abstract}
\tolerance=3000%
Для двопiдґраткової моделi жорстких бозонiв в рамках наближення хаотичних фаз розраховано
енергетичний спектр i спектральнi густини у рiзних фазах та побудовано фазовi дiаграми.
Дослiджено перебудову бозонного спектру при змiнi температури, хiмiчного потенцiалу та рiзницi
енергiй локальних позицiй у пiдґратках. Побудовано фазовi дiаграми, якi iлюструють областi
iснування нормальної фази, що може бути подiбною до фази моттiвського дiелектрика (MI) чи
зарядового впорядкування (CDW), а також фази з бозе-конденсатом (фази SF).

\keywords жорсткi бозони, густина станiв, фазовi дiаграми
\end{abstract}

\end{document}